\documentclass[a4paper]{article}
\usepackage[a4paper,hmargin=1.5cm]{geometry}
\usepackage[toc,page]{appendix}

\usepackage{times}
\usepackage[T1]{fontenc}
\usepackage[utf8]{inputenc}
\usepackage{array}
\usepackage{multirow}
\usepackage{float}
\usepackage{url}
\usepackage{amsthm}
\usepackage{amsmath}
\usepackage{amssymb}
\usepackage{graphicx}
\usepackage{color}
\usepackage{caption}
\usepackage[english]{babel}
\usepackage{dsfont} 
\usepackage[numbered,framed]{mcode}

\renewcommand{\L}{\mathcal{L}} 
 
\newcommand{\W}{\mathbf{W}}

\newcommand{\mc}[1]{{\bf{[M]: }{{\ttfamily \hyphenchar\the\font=`_}\texttt{\detokenize{#1}}}}}
\newcommand{\pc}[1]{{\bf{[P]: }{{\ttfamily \hyphenchar\the\font=`.}{\ttfamily \hyphenchar\the\font=`_}\texttt{\detokenize{#1}}}}}

\begin{document}

\title{GSPBOX: A toolbox for signal processing on graphs}

\author{Nathanael Perraudin, Johan Paratte, David Shuman, Lionel Martin\\ Vassilis Kalofolias, Pierre Vandergheynst and David K. Hammond}

\maketitle

\begin{abstract}
This document introduces the Graph Signal Processing Toolbox (GSPBox) a framework that can be used to tackle graph related problems with a signal processing approach. It explains the structure and the organization of this software. It also contains a general description of the important modules. 
\end{abstract}


\section{Toolbox organization}

In this document, we briefly describe the different modules available in the toolbox. For each of them, the main functions are briefly described. This chapter should help making the connection between the theoretical concepts introduced in \cite{shuman2013emerging,shuman2013vertex,shuman2013multiscale} and the technical documentation provided with the toolbox. We highly recommend to read this document and the tutorial before using the toolbox. The documentation, the tutorials and other resources are available on-line\footnote{See \url{https://lts2.epfl.ch/gsp/doc/} for MATLAB and \url{https://lts2.epfl.ch/pygsp} for Python. The full documentation is also available in a single document: \url{https://lts2.epfl.ch/gsp/gspbox.pdf}}. 

The toolbox has first been implemented in MATLAB but a port to Python, called the PyGSP, has been made recently. As of the time of writing of this document, not all the functionalities have been ported to Python, but the main modules are already available. In the following, functions prefixed by \mc{} refer to the MATLAB implementation and the ones prefixed with \pc{} refer to the Python implementation. 

\subsection{General structure of the toolbox (MATLAB)}
The general design of the GSPBox focuses around the graph object \cite{shuman2013emerging}, a MATLAB structure containing the necessary informations to use most of the algorithms. By default, only a few attributes are available (see section \ref{sec:toolbox_graph}), allowing only the use of a subset of functions. In order to enable the use of more algorithms, additional fields can be added to the graph structure. For example, the following line will compute the graph Fourier basis enabling exact filtering operations.
\begin{lstlisting}
    G = gsp_compute_fourier_basis(G);
\end{lstlisting}
Ideally, this operation should be done on the fly when exact filtering is required. Unfortunately, the lack of well defined class paradigm in MATLAB makes it too complicated to be implemented. Luckily, the above formulation prevents any unnecessary data copy of the data contained in the structure \texttt{G}. In order to avoid name conflicts, all functions in the GSPBox start with \mc{gsp_}.
A second important convention is that all functions applying a graph algorithm on a graph signal takes the graph as first argument. For example, the graph Fourier transform of the vector \texttt{f} is computed by 
\begin{lstlisting}
    fhat = gsp_gft(G,f);
\end{lstlisting}
The graph operators are described in section \ref{sec:toolbox_op}. Filtering a signal on a graph is also a linear operation. However, since the design of special filters (kernels) is important, they are regrouped in a dedicated module (see section \ref{sec:toolbox_filters}). 

The toolbox contains two additional important modules. The optimization module contains proximal operators, projections and solvers compatible with the UNLocBoX \cite{perraudin2014unlocbox} (see section \ref{sec:toolbox_opt}). These functions facilitate the definition of convex optimization problems using graphs. Finally, section \ref{sec:toolbox_ml} is composed of well known graph machine learning algorithms.

\subsection{General structure of the toolbox (Python)}

The structure of the Python toolbox follows closely the MATLAB one. The major difference comes from the fact that the Python implementation is object-oriented and thus allows for a natural use of instances of the graph object. 

For example the equivalent of the MATLAB call:
\begin{lstlisting}
    G = gsp_estimate_lmax(G);
\end{lstlisting}
can be achieved using a simple method call on the graph object:
\begin{lstlisting}
    G.estimate_lmax()
\end{lstlisting}

Moreover, the use of class for the "graph object" allows to compute additional graph attributes on the fly, making the code clearer as its MATLAB equivalent.
Note though that functionalities are grouped into different modules (one per section below) and that several functions that work on graphs have to be called directly from the modules. For example, one should write:
\begin{lstlisting}
    layers = pygsp.operators.kron_pyramid(G, levels)
\end{lstlisting}

This is the case as soon as the graph is the structure on which the action has to be performed and not our principal focus.

In a similar way to the MATLAB implementation using the UNLocBoX for the convex optimization routines, the Python implementation uses the PyUNLocBoX, which is the Python port of the UNLocBoX.

\section{Graphs} \label{sec:toolbox_graph}
The GSPBox is constructed around one main object: the graph. It is implemented as a structure in Matlab and as a class in Python. It stores the nodes, the edges and other attributes related to the graph. In the implementation, a graph is fully defined by the weight matrix $\W$, which is the main and only required attribute. Since most graph structures are far from fully connected, $\W$ is implemented as a sparse matrix. From the weight matrix a Laplacian matrix $\L$ is computed and stored as an attribute of the graph object. Different other attributes are available such as plotting attributes, vertex coordinates, the degree matrix, the number of vertices and edges. The list of all attributes is given in table \ref{table:graph_attr}.

\begin{center}
    \begin{tabular}{ | l | l | l | l |}
    \hline
    Attribute & Format & Data type & Description \\ \hline
    \hline
    \multicolumn{4}{|c|}{ \textbf{Mandatory fields} }  \\ \hline
    \texttt{W} & $N$x$N$ sparse matrix & double & Weight matrix $\W$ \\ \hline
    \texttt{L} & $N$x$N$ sparse matrix & double & Laplacian matrix $\L$ \\ \hline    
   \texttt{d} & $N$x$1$ vector & double & The diagonal of the degree matrix \\ \hline
   \texttt{N} & scalar & integer & Number of vertices \\ \hline    
   \texttt{Ne} & scalar & integer & Number of edges \\ \hline 
    \texttt{plotting} & \mc{structure} \pc{dict} & none & Plotting parameters \\ \hline
	\texttt{type} & text & string & Name, type or short description \\ \hline    
	\texttt{directed} & scalar & \mc{logical} \pc{boolean} & State if the graph is directed or not \\ \hline
   \texttt{lap\_type} & text & string & Laplacian type \\ \hline
    \hline
    \multicolumn{4}{|c|}{ \textbf{Optional fields} }  \\
  \hline
      \texttt{A} & $N$x$N$ sparse matrix & \mc{logical} \pc{boolean} & Adjacency matrix \\ \hline        
    \texttt{coords} & $N$x$2$ or $N$x$3$ matrix & double & Vectors of coordinates in 2D or 3D. \\ \hline
    \texttt{lmax} & scalar & double & Exact or estimated maximum eigenvalue \\ 
    \hline
    \texttt{U} & $N$x$N$ matrix & double & Matrix of eigenvectors \\ \hline 
    \texttt{e} & $N$x$1$ vector & double & Vector of eigenvalues \\ \hline 
    \texttt{mu} & scalar & double & Graph coherence \\ \hline 
    \end{tabular}
    \captionof{table}{Attributes of the graph object}
    \label{table:graph_attr}
\end{center}

The easiest way to create a graph is the \mc{gsp_graph} \pc{pygsp.graphs.Graph} function which takes the weight matrix as input. This function initializes a graph structure by creating the graph Laplacian and other useful attributes. Note that by default the toolbox uses the combinatorial definition of the Laplacian operator. Other Laplacians can be computed using the \mc{gsp_create_laplacian} \pc{pygsp.gutils.create_laplacian} function. Please note that almost all functions are dependent of the Laplacian definition. As a result, it is important to select the correct definition at first.

Many particular graphs are also available using helper functions such as : ring, path, comet, swiss roll, airfoil or two moons. In addition, functions are provided for usual non-deterministic graphs such as : Erdos-Renyi, community, Stochastic Block Model or sensor networks graphs.

Nearest Neighbors (NN) graphs form a class which is used in many applications and can be constructed from a set of points (or point cloud) using the \mc{gsp_nn_graph} \pc{pygsp.graphs.NNGraph} function. The function is highly tunable and can handle very large sets of points using FLANN \cite{flann_pami_2014}.

Two particular cases of NN graphs have their dedicated helper functions : 3D point clouds and image patch-graphs. An example of the former can be seen in the function \mc{gsp_bunny} \pc{pygsp.graphs.Bunny}. As for the second, a graph can be created from an image by connecting similar patches of pixels together. The function \mc{gsp_patch_graph} creates this graph. Parameters allow the resulting graph to vary between local and non-local and to use different distance functions \cite{zhang2008graph,narang2012graph}. 

\begin{figure}[htb!]
\begin{center}
\includegraphics[width=0.45\textwidth]{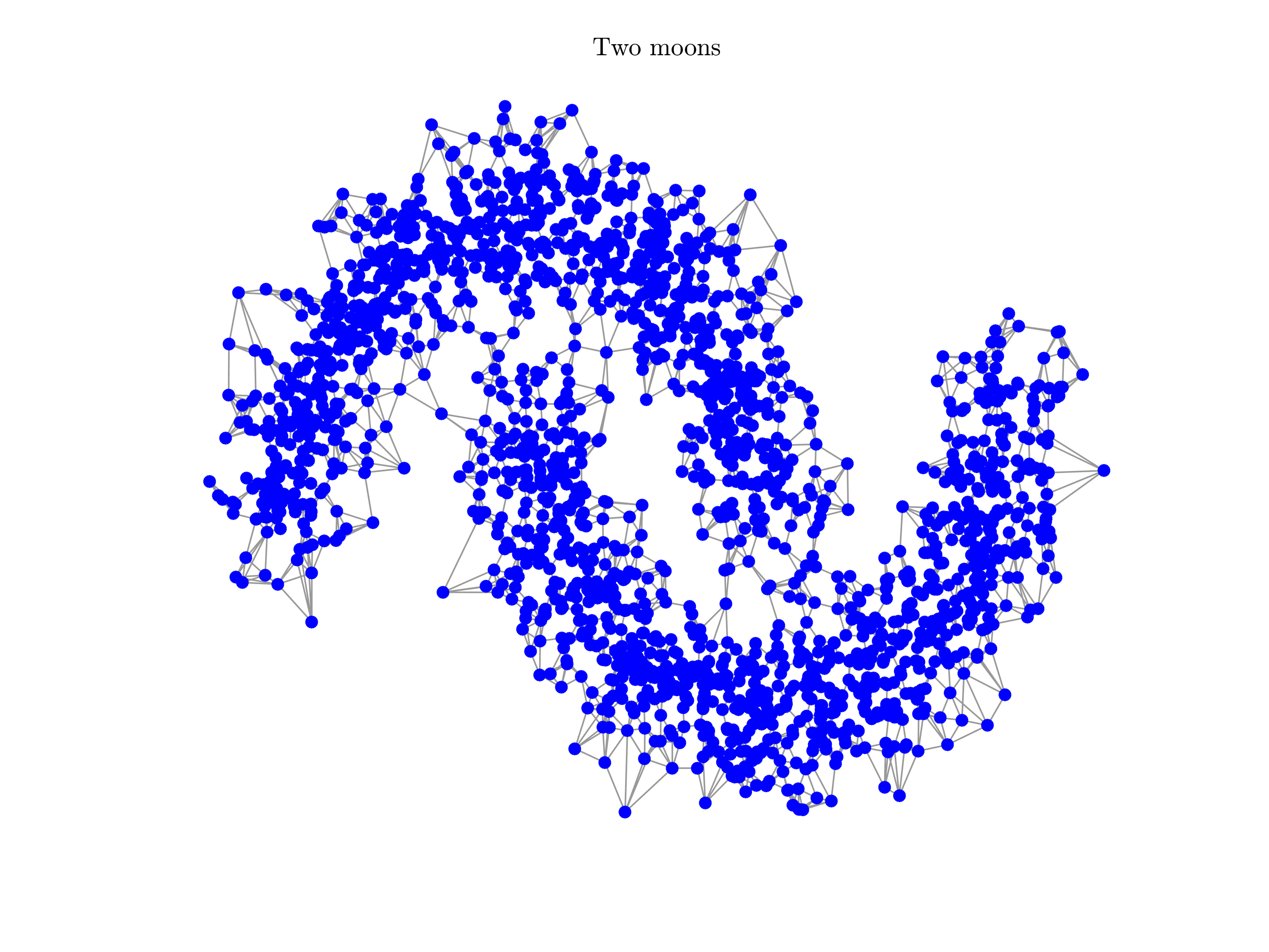} 
\includegraphics[width=0.45\textwidth]{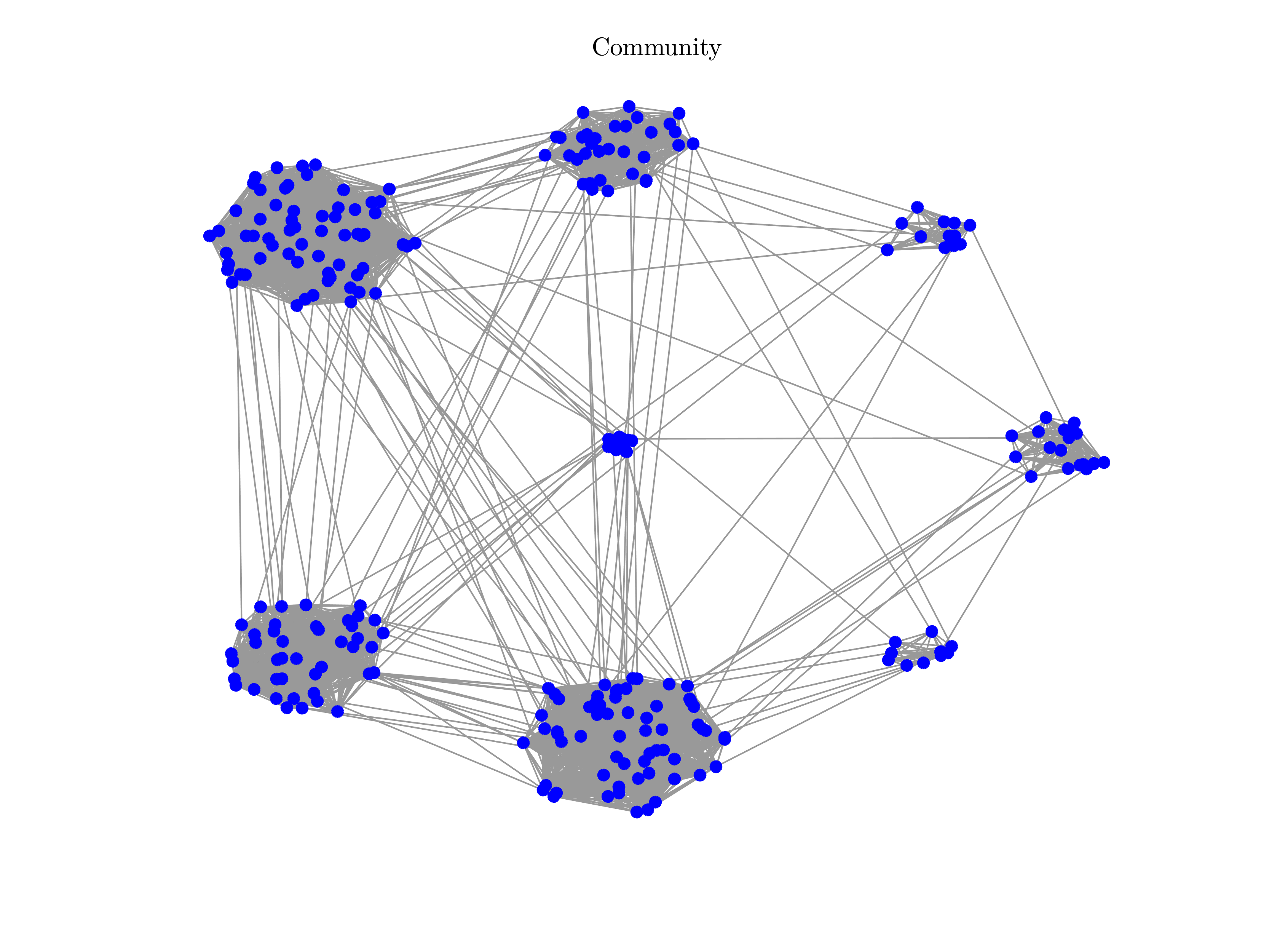} 
\includegraphics[width=0.45\textwidth]{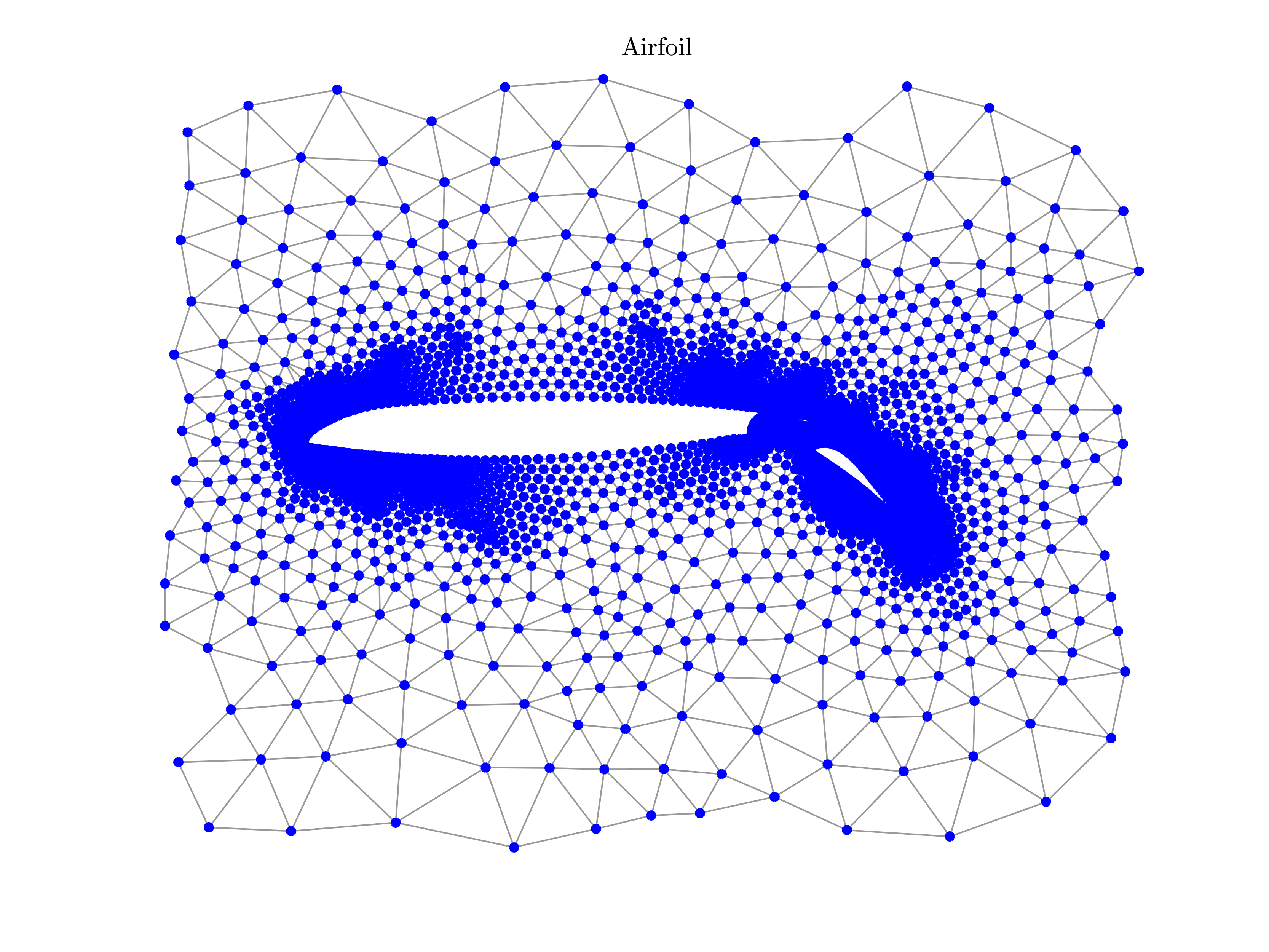}
\includegraphics[width=0.45\textwidth]{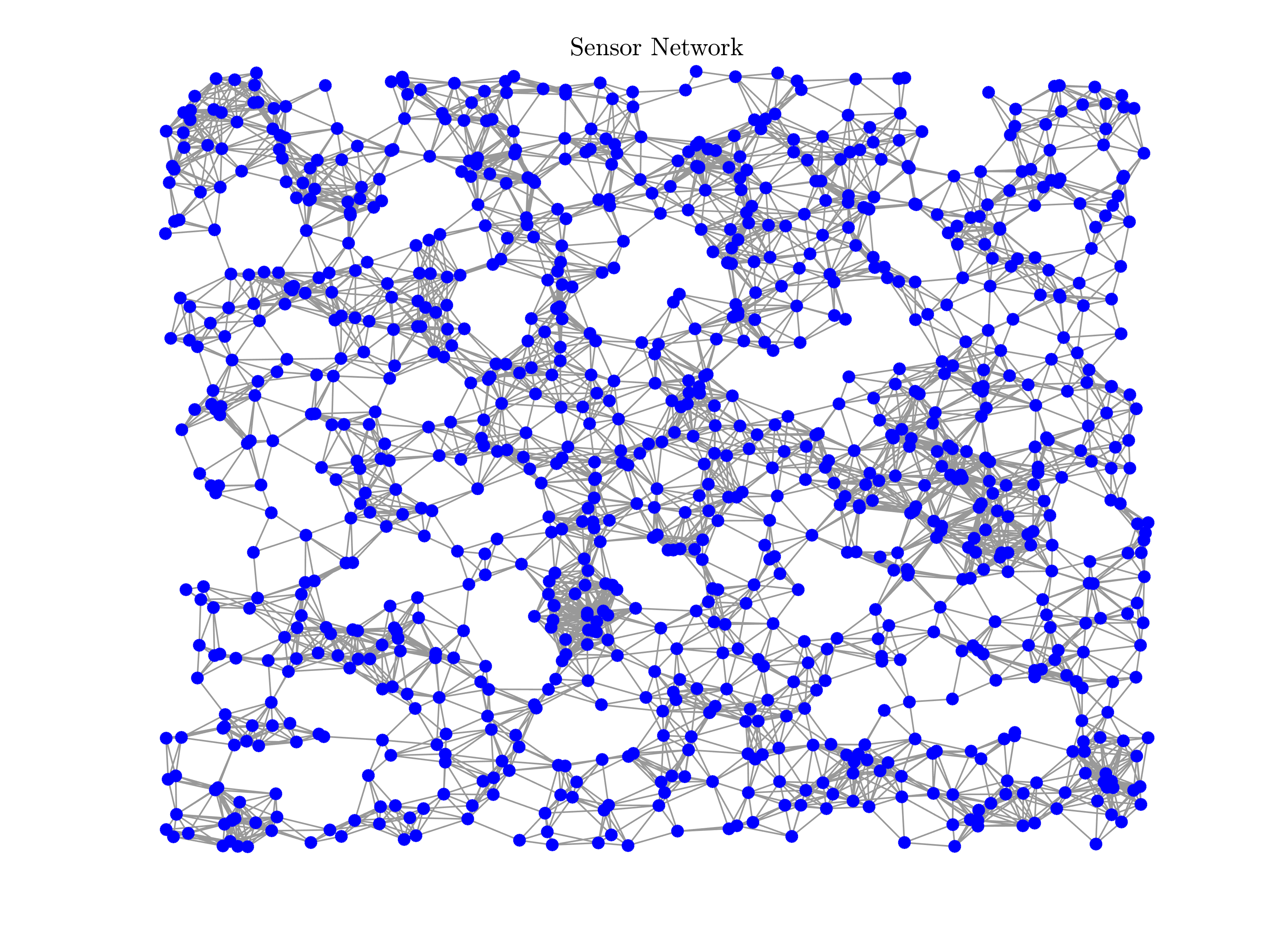}
\end{center}
\caption{Examples of classical graphs : two moons (top left), community (top right), airfoil (bottom left) and sensor network (bottom right).}
\label{fig:toolbox_graphs}
\end{figure}

A few examples of the graphs are displayed in Figure\ref{fig:toolbox_graphs}. 

\section{Plotting}

As in many other domains, visualization is very important in graph signal processing. The most basic operation is to visualize graphs. This can be achieved using a call to the function \mc{gsp_plot_graph} \pc{pygsp.plotting.plot_graph}. In order to be displayable, a graph needs to have 2D (or 3D) coordinates (which is a field of the graph object). Some graphs do not possess default coordinates (e.g. Erdos-Renyi).

The toolbox also contains routines to plot signals living on graphs. The function dedicated to this task is \mc{gsp_plot_signal} \pc{pygsp.plotting.plot_signal}. For now, only 1D signals are supported. By default, the value of the signal is displayed using a color coding, but bars can be displayed by passing parameters. 

The third visualization helper is a function to plot filters (in the spectral domain) which is called \mc{gsp_plot_filter} \pc{pygsp.plotting.plot_filter}. It also supports filter-banks and allows to automatically inspect the related frames.

\begin{figure}[htb!]
\begin{center}
\includegraphics[width=0.225\textwidth]{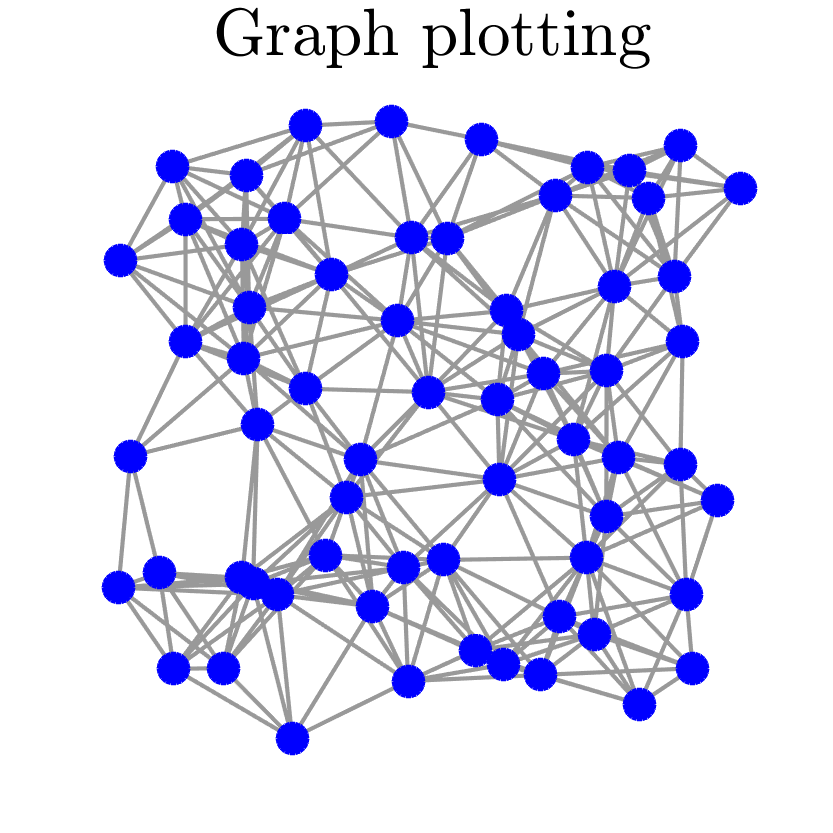} 
\includegraphics[width=0.225\textwidth]{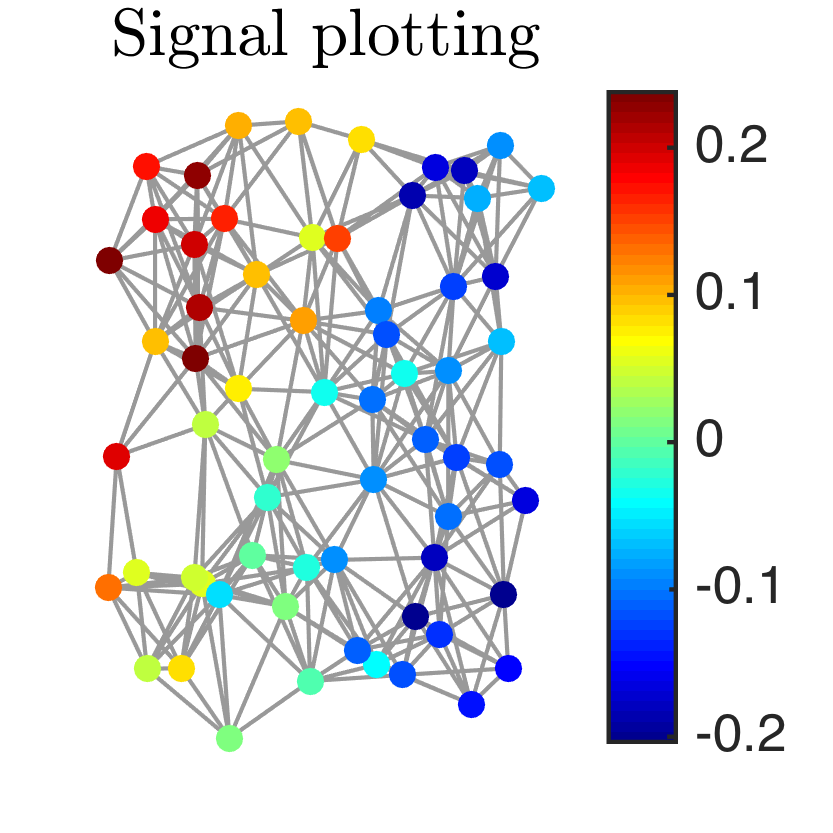} 
\includegraphics[width=0.225\textwidth]{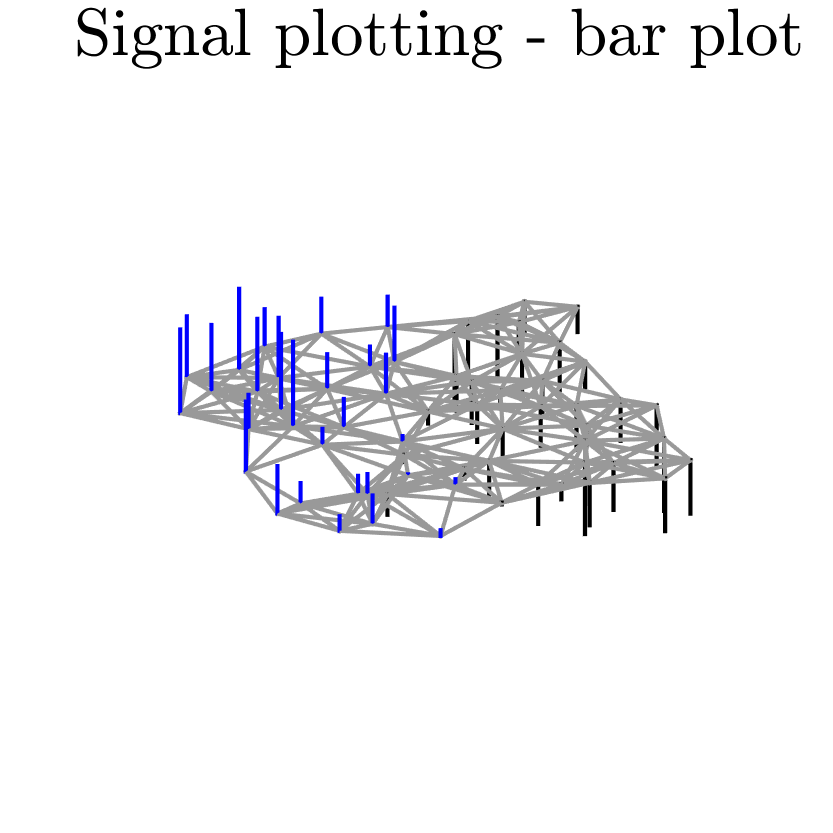} 
\includegraphics[width=0.225\textwidth]{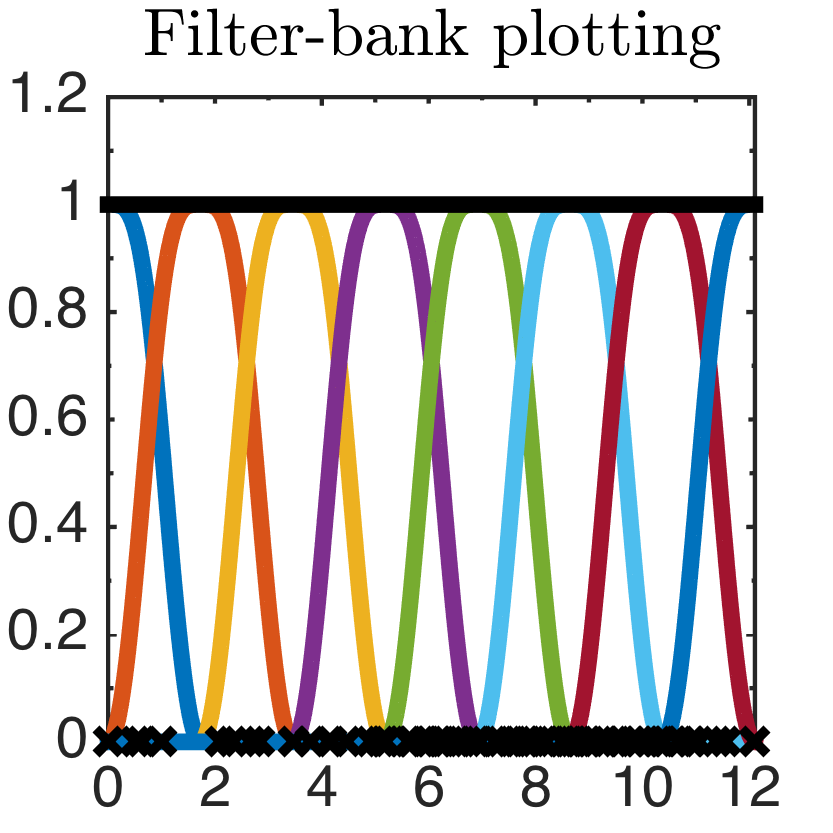}
\end{center}
\caption{Visualization of graph and signals using plotting functions.}
\label{fig:toolbox_plotting}
\end{figure}

The results obtained using these three plotting functions are visible in Fig.~\ref{fig:toolbox_plotting}. 

\section{Operators} \label{sec:toolbox_op}
The module operators contains basics spectral graph functions such as Fourier transform, localization, gradient, divergence or pyramid decomposition. Since all operator are based on the Laplacian definition, the necessary underlying objects (attributes) are all stored into a single object: the graph. 

As a first example, the graph Fourier transform \mc{gsp_gft} \pc{pygsp.operators.gft} requires the Fourier basis. This attribute can be computed with the function \mc{gsp_compute_fourier_basis} \pc{pygsp.graphs.compute_fourier_basis} \cite{shuman2013vertex} that adds the fields \texttt{U}, \texttt{e} and \texttt{lmax} to the graph structure. As a second example, since the gradient and divergence operate on the edges of the graph, a search on the edge matrix is needed to enable the use of these operators. It can be done with the routines \mc{gsp_adj2vec} \pc{pygsp.operators.adj2vec}. These operations take time and should be performed only once. In MATLAB, these functions are called explicitly by the user beforehand. However, in Python they are automatically called when needed and the result stored as an attribute.

The module operator also includes a Multi-scale Pyramid Transform for graph signals \cite{shuman2013multiscale}. Again, it works in two steps. First the pyramid is precomputed with \mc{gsp_graph_multiresolution} \pc{pygsp.operators.graph_multiresolution}. Second the decomposition of a signal is performed with \mc{gsp_pyramid_analysis} \pc{pygsp.operators.pyramid_analysis}. The reconstruction uses \mc{gsp_pyramid_synthesis} \pc{pygsp.operators.pyramid_synthesis}.

The Laplacian is a special operator stored as a sparse matrix in the field \texttt{L} of the graph. Table \ref{tab:deflap} summarizes the available definitions. We are planning to implement additional ones.
\begin{table}
\begin{center}

\begin{tabular}{|l|l|l|l|}
  \hline
  Name & Edge derivative  $\frac{\partial f}{\partial_e}(i,j)$& Laplacian matrix (operator)& Available\\
  \hline
  \hline
\multicolumn{4}{|c|}{ \textbf{Undirected graph} }  \\
 \hline
Combinatorial Laplacian & $\sqrt{\W(i,j)} \left( f(j)-f(i) \right)$ & $ \mathbf{D}-\W $ & \bf{V} \\
  \hline
  Normalized Laplacian  & $\sqrt{\W(i,j)}\left(\frac{f(j)}{\sqrt{d(j)}}-\frac{f(i)}{\sqrt{d(i)}}\right)$ & $\mathbf{D}^{-\frac{1}{2}} (\mathbf{D}-\W) \mathbf{D}^{-\frac{1}{2}}$ & \bf{V} \\
  \hline
    \hline
\multicolumn{4}{|c|}{ \textbf{Directed graph} }  \\
  \hline
  Combinatorial Laplacian & $\sqrt{\W(i,j)} \left( f(j)-f(i) \right)$  & $\frac{1}{2}\left(\mathbf{D}_{+}+\mathbf{D}_{-}-\mathbf{W}-\mathbf{W}^{*}\right)$ & \bf{V}\\
\hline
 Degree normalized Laplacian & $\sqrt{\W(i,j)}\left(\frac{f(j)}{\sqrt{d_-(j)}}-\frac{f(i)}{\sqrt{d_+(i)}}\right)$ & $\mathbf{I}-\frac{1}{2}\left(\mathbf{D}_{+}^{-\frac{1}{2}}[\mathbf{W}+\mathbf{W}^{*}]\mathbf{D}_{-}^{-\frac{1}{2}}\right)$ & \bf{V}\\  
    \hline
   Distribution normalized Laplacian & $\sqrt{\pi(i)}\left(\sqrt{\frac{p(i,j)}{\pi(j)}}f(j)-\sqrt{\frac{p(i,j)}{\pi(i)}}f(i)\right)$ & $\frac{1}{2}\left(\mathbf{\Pi}^{\frac{1}{2}}\mathbf{P}\mathbf{\Pi}^{-\frac{1}{2}}+\mathbf{\Pi}^{-\frac{1}{2}}\mathbf{P}^{*}\mathbf{\Pi}^{\frac{1}{2}}\right)$ & \bf{V} \\   
 \hline   

\end{tabular}
\caption{\label{tab:deflap} Different definitions for graph Laplacian operator and their associated edge derivative. (For directed graph, $d_+,\mathbf{D}_+$  and $d_-,\mathbf{D}_-$ define the out degree and in-degree of a node. $\pi,\mathbf{\Pi}$ is the stationary distribution of the graph and $\mathbf{P}$ is a normalized weight matrix $\mathbf{W}$. For sake of clarity, exact definition of those quantities are not given here, but can be found in \cite{zhou2005learning}.)}
\end{center}
\end{table}

\section{Filters} \label{sec:toolbox_filters}
Filters are a special kind of linear operators that are so prominent in the toolbox that they deserve their own module \cite{shuman2013vertex,shuman2013emerging,hammond2011wavelets,shuman2012windowed,hammond2011wavelets}. A filter is simply an anonymous function (in MATLAB) or a lambda function (in Python) acting element-by-element on the input. In MATLAB, a filter-bank is created simply by gathering these functions together into a cell array. For example, you would write:
\newpage
\begin{lstlisting}
    % g(x) = x^2 + sin(x)
    g = @(x) x.^2 + sin(x);
    % h(x) = exp(-x)
    h = @(x) exp(-x);
    % Filterbank composed of g and h
    fb = {g,h};
\end{lstlisting}

The toolbox contains many predefined design of filter. They all start with \mc{gsp_design_} in MATLAB and are in the module \pc{pygsp.filters} in Python. Once a filter (or a filter-bank) is created, it can be applied to a signal with \mc{gsp_filter_analysis} in MATLAB and a call to the method \pc{analysis} of the filter object in Python. Note that the toolbox uses accelerated algorithms to scale almost linearly with the number of sample \cite{susnjara2015accelerated}. 

The available type of filter design of the GSPBox can be classified as:
\begin{itemize}
	\item Wavelets (Filters are scaled version of a mother window)
	\item Gabor (Filters are shifted version of a mother window)
	\item Low pass filter (Filters to de-noise a signal)
	\item High pass / Low pass separation filterbank (tight frame of 2 filters to separate the high frequencies from the low ones. No energy is lost in the process)
\end{itemize}
Additionally, to adapt the filter to the graph eigen-distribution, the warping function \mc{gsp_design_warped_translates} \pc{pygsp.filters.WarpedTranslates} can be used~\cite{shuman2013spectrum}.

\section{UNLocBoX Binding}\label{sec:toolbox_opt}
This module contains special wrappers for the UNLocBoX\cite{perraudin2014unlocbox}. It allows to solve convex problems containing graph terms very easily \cite{zhou2004learning,zhou2004regularization,zhou2005learning,belkin2006manifold}. For example, the proximal operator of the graph TV norm is given by \mc{gsp_prox_tv}. The optimization module contains also some predefined problems such as graph basis pursuit in \mc{gsp_solve_l1} or wavelet de-noising in \mc{gsp_wavelet_dn}. There is still active work on this module so it is expected to grow rapidly in the future releases of the toolbox.



\section{Toolbox conventions}

\subsection{General conventions}

\begin{itemize}
\item As much as possible, all small letters are used for vectors (or vector stacked into a matrix) and capital are reserved for matrices. A notable exception is the creation of nearest neighbors graphs. 
	\item A variable should never have the same name as an already existing function in MATLAB or Python respectively. This makes the code easier to read and less prone to errors. This is a best coding practice in general, but since both languages allow the override of built-in functions, a special care is needed.
	\item All function names should be lowercase. This avoids a lot of confusion because some computer architectures respect upper/lower casing and others do not. 
	\item As much as possible, functions are named after the action they perform, rather than the algorithm they use, or the person who invented it.
	\item No global variables. Global variables makes it harder to debug and the code is harder to parallelize.
\end{itemize}

\subsection{MATLAB}
\begin{itemize}
	\item All function start by \texttt{gsp\_}.
	\item The graph structure is always the first argument in the function call. Filters are always second. Finally, optional parameter are last. 
	\item In the toolbox, we do use any argument helper functions. As a result, optional argument are generally stacked into a graph structure named \texttt{param}.
	\item If a transform works on a matrix, it will per default work along the columns. This is a standard in Matlab (fft does this, among many other functions).
	\item Function names are traditionally written in uppercase in MATLAB documentation.
\end{itemize}

\subsection{Python}
\begin{itemize}
	\item All functions should be part of a module, there should be no call directly from pygsp (\pc{pygsp.my_function}).
	\item Inside a given module, functionalities can be further split in different files regrouping those that are used in the same context.
	\item MATLAB's matrix operations are sometimes ported in a different way that preserves the efficiency of the code. When matrix operations are necessary, they are all performed through the \texttt{numpy} and \texttt{scipy} libraries.
	\item Since Python does not come with a plotting library, we support both \texttt{matplotlib} and \texttt{pyqtgraph}. One should install the required libraries on his own. If both are correctly installed, then \texttt{pyqtgraph} is favoured unless specifically specified.
\end{itemize}

\section*{Acknowledgements}
We would like to thanks all coding authors of the GSPBOX. 
The toolbox was ported in Python by Basile Châtillon, Alexandre Lafaye and Nicolas Rod. The toolbox was also improved by Nauman Shahid and Yann Schönenberger.

{\small \bibliographystyle{abbrv}
\bibliography{biblio}
}{\small \par}

\end{document}